\documentclass[twocolumn,secnumarabic,amssymb, nobibnotes, aps, prl, showpacs, superscriptaddress]{revtex4}

\usepackage{graphicx}
\usepackage{dcolumn}
\usepackage{bm}
\usepackage{color} 

\begin{document}


\title{Anisotropic and strong negative magneto-resistance in the three-dimensional topological insulator Bi$_2$Se$_3$}

\author{S.~Wiedmann}
\email{s.wiedmann@science.ru.nl}
\affiliation{High Field Magnet Laboratory (HFML-EMFL) \& Institute for Molecules and Materials, Radboud University, Toernooiveld 7, 6525 ED Nijmegen, The Netherlands}

\author{A.~Jost}
\affiliation{High Field Magnet Laboratory (HFML-EMFL) \& Institute for Molecules and Materials, Radboud University, Toernooiveld 7, 6525 ED Nijmegen, The Netherlands}

\author{B.~Fauqu\'{e}}
\affiliation{LPEM (CNRS-UPMC), ESPCI, 75005 Paris, France}

\author{J.~van~Dijk}
\affiliation{High Field Magnet Laboratory (HFML-EMFL) \& Institute for Molecules and Materials, Radboud University, Toernooiveld 7, 6525 ED Nijmegen, The Netherlands}

\author{M.~J.~Meijer}
\affiliation{High Field Magnet Laboratory (HFML-EMFL) \& Institute for Molecules and Materials, Radboud University, Toernooiveld 7, 6525 ED Nijmegen, The Netherlands}

\author{T.~Khouri}
\affiliation{High Field Magnet Laboratory (HFML-EMFL) \& Institute for Molecules and Materials, Radboud University, Toernooiveld 7, 6525 ED Nijmegen, The Netherlands}

\author{S.~Pezzini}
\affiliation{High Field Magnet Laboratory (HFML-EMFL) \& Institute for Molecules and Materials, Radboud University, Toernooiveld 7, 6525 ED Nijmegen, The Netherlands}

\author{S.~Grauer}
\affiliation{Physikalisches Institut (EP3), Universit\"at W\"urzburg, Am Hubland, 97074 W\"urzburg,Germany}

\author{S.~Schreyeck}
\affiliation{Physikalisches Institut (EP3), Universit\"at W\"urzburg, Am Hubland, 97074 W\"urzburg,Germany}

\author{C.~Br\"une}
\affiliation{Physikalisches Institut (EP3), Universit\"at W\"urzburg, Am Hubland, 97074 W\"urzburg,Germany}

\author{H.~Buhmann}
\affiliation{Physikalisches Institut (EP3), Universit\"at W\"urzburg, Am Hubland, 97074 W\"urzburg,Germany}

\author{L.~W.~Molenkamp}
\affiliation{Physikalisches Institut (EP3), Universit\"at W\"urzburg, Am Hubland, 97074 W\"urzburg,Germany}

\author{N.~E.~Hussey}
\affiliation{High Field Magnet Laboratory (HFML-EMFL) \& Institute for Molecules and Materials, Radboud University, Toernooiveld 7, 6525 ED Nijmegen, The Netherlands}

\date{\today}

\begin{abstract}
We report on high-field angle-dependent magneto-transport measurements on epitaxial thin films of Bi$_2$Se$_3$, 
a three-dimensional topological insulator. At low temperature, we observe  quantum oscillations that demonstrate
the simultaneous presence of bulk and surface carriers. The magneto-resistance of Bi$_2$Se$_3$ is found to be 
highly anisotropic. In the presence of a parallel electric and magnetic field, we observe a strong negative 
longitudinal magneto-resistance that has been considered as a smoking-gun for the presence of chiral 
fermions in a certain class of semi-metals due to the so-called axial anomaly. Its observation in a 
three-dimensional topological insulator implies that the axial anomaly may be in fact a far more generic phenomenon than 
originally thought.
\end{abstract}


\maketitle


The role of topology in condensed matter systems, once a rather esoteric pursuit, has undergone a revolution
in the last decade with the realization that a certain class of insulators and semi-metals play host to 
topologically-protected surface states. In 2009, band structure calculations revealed that stoichiometric
Bi$_2$Se$_3$, a well-known thermoelectric material \cite{1}, bears all the hallmarks of a three-dimensional 
topological insulator (3D TI) \cite{2} with an insulating bulk and conducting surface states provided 
that the Fermi energy $\epsilon_F$ is situated within the bulk band gap \cite{3}. These gapless surface 
states possess opposite spin and momentum, and are protected from backscattering by time reversal symmetry. 
The existence of Dirac-like surface states within the bulk band gap was confirmed in an angle-resolved 
photoemission spectroscopy study performed that same year \cite{4}.

Though Bi$_2$Se$_3$ is arguably the most simple representative of the 3D TI family,
accessing the topological surface states (TSS) in transport has been hindered by a large
residual carrier density in the bulk\cite{5,6}. While Shubnikov-de Haas (SdH)
oscillations are a powerful means to distinguish between bulk and surface charge
carriers via their angle dependence, their analysis and interpretation remain controversial.
The literature is replete with results that have been attributed to single-bands of bulk
carriers, TSS or multiple bands, emphasizing the difficulty in distinguishing between bulk,
TSS and a two-dimensional charge-accumulation layer \cite{5,6,7,8,9,10,11}. Apart from
the TSS, the electronic bulk states in Bi$_2$Se$_3$ are of particular interest
since their spin splitting is found to be twice the cyclotron energy observed in quantum
oscillation \cite{12,13} and optical \cite{14} experiments. Another peculiar property of Bi$_2$Se$_3$ 
and other 3D TIs is the observation of a linear positive magneto-resistance (MR) that persists 
up to room temperature \cite{15,16,17,18,19,20}.

The recent explosion of interest in 3D massless Dirac fermions in `3D Dirac' or `Weyl'
semi-metals \cite{21} is based primarily on their unique topological properties that can be
revealed in relatively straightforward magneto-transport experiments. Examples
include the observation of an extremely large positive MR \cite{22}, linear MR \cite{23}
and, more specifically, the negative longitudinal MR (NLMR) predicted to appear in Weyl semi-metals
when the magnetic and electric field are co-aligned. This NLMR has been attributed
to the axial anomaly, a quantum mechanical phenomenon that relies on a number
imbalance of chiral fermions in the presence of an applied electric field
\cite{22,24,25,26}. In a recent theoretical study, however, it was proposed that
the NLMR phenomenon may in fact be a generic property of metals and semiconductors \cite{27},
rather than something unique to topological semi-metals.

In this Rapid Communication, we present magneto-transport experiments on Bi$_2$Se$_3$ epitaxial layers
in magnetic fields up to 30~T. At low-temperatures, we establish the existence
of both bulk and surface carriers via angle-dependent SdH measurements. Moreover,
we observe a strong anisotropy in the MR which depends on the orientation of the current
$I$ with respect to the applied magnetic field $B$ over a wide range of carrier concentrations. 
When the magnetic field is applied parallel to $I$ ($I \parallel B_x$), we observe a strong 
NLMR. This surprising finding confirms that the observation of NLMR is not unique to Weyl 
semi-metals and therefore cannot by itself be taken as conclusive evidence for the existence 
of Weyl fermions in other systems. With this in mind, we consider possible alternative origins 
of this increasingly ubiquitous phenomenon, but argue finally that the axial anomaly may indeed be
generic to a host of three-dimensional materials \cite{27}.

The present study has been performed on samples with different layer thicknesses 
$d$=290, 190, 50 and 20~nm (referred to hereafter as samples \#A, \#B, \#C and \#D) grown by molecular 
beam epitaxy (MBE) on an InP(111)B substrate  \cite{28} and patterned in a six-terminal Hall-bar geometry 
(length $L~\times$~width $W$ - (30$\times$10)$~\mu$m$^2$). 
The carrier concentration $n = n_{Hall}$ at 300~K (extracted from the linear part of the low-field Hall resistivity 
$\rho_{xy}$) varies from 1.2$\cdot$10$^{18}$-1.7$\cdot$10$^{19}$cm$^{-3}$ with decreasing 
thickness \cite{29}. All magneto-transport measurements reported here were performed in a $^4$He flow 
cryostat in a resistive (Bitter) magnet up to 30~T using standard ac lock-in detection techniques with 
an excitation current of 1$\mu$A. For the SdH oscillation analysis, the magnetic field is applied in a plane 
perpendicular to the current $I$.

\begin{figure}[ht]
\includegraphics[width=8cm]{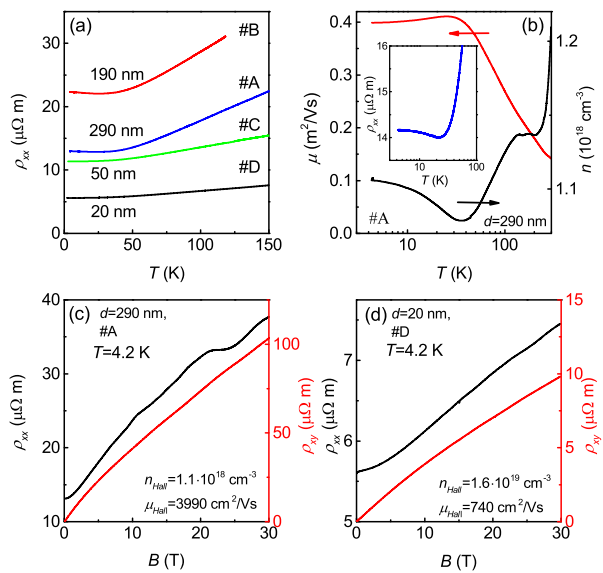}
\caption{\label{fig1} (Color online) (a) Temperature dependence of the electrical resistivity $\rho_{xx}$ for Bi$_2$Se$_3$
MBE-grown films with different thicknesses. (b) Mobility $\mu$ and carrier density $n$ as
a function of $T$ for sample \#A. (c,d) $\rho_{xx}$ and $\rho_{xy}$ as a function of $B$ at $T$=4.2~K for
\#A and \#D, respectively. The quoted carrier density is extracted from the low-field Hall resistance.}
\end{figure}

The temperature dependence of the longitudinal resistivity $\rho_{xx}$ is shown in
Fig.~\ref{fig1}(a) for all four samples. In Fig.~\ref{fig1}(b), we plot the carrier
mobility $\mu$ and concentration $n=1/(R_{H}e)$ for \#A obtained from the zero-field $\rho_{xx}(T)$
sweep and the measured $\rho_{xy}$ at $B$=1~T, respectively. The overall temperature
dependence is metallic ($dR_{xx}/dT>$~0) though below 40~K, we observe a tiny upturn
in $\rho_{xx}$ which is strongest for the sample with the lowest carrier density. This
increase is accompanied by a small decrease in $\mu$ and an apparent increase in $n$ which 
has been interpreted to originate from the presence of an impurity band \cite{7,8,30}. In Figs.~\ref{fig1}(c) 
and (d), we plot $\rho_{xx}$ and $\rho_{xy}$ as a function of the magnetic field $B$ up to 30~T 
for samples \#A and \#D. SdH oscillations are superimposed on top of a positive  quasi-linear MR 
while $\rho_{xy}$ is found to be non-linear for $B>$ 2~T suggesting the possible presence of two 
carrier types. From the low-field $\rho_{xy}$, we extract a carrier mobility of 3990 (740)~cm$^{2}/$Vs 
for sample \#A (\#D) respectively.

\begin{figure}[ht]
\includegraphics[width=8cm]{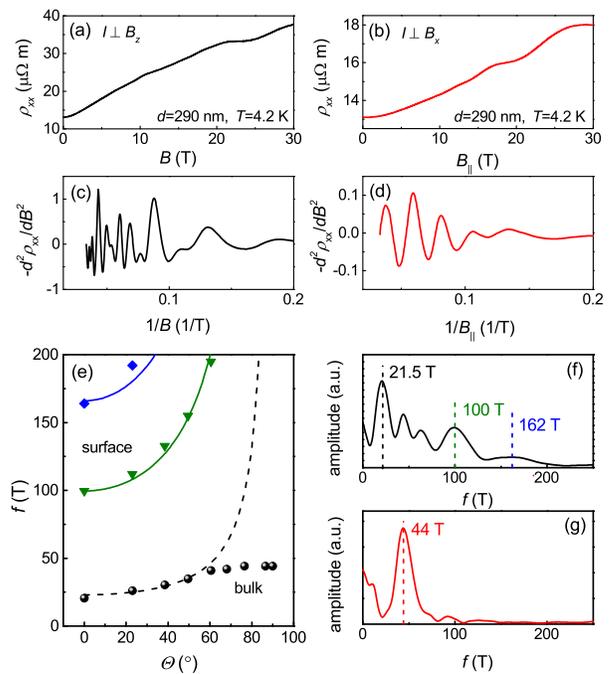}
\caption{\label{fig2} (Color online) Longitudinal resistivity $\rho_{xx}$ as a function
of (a) perpendicular and (b) in-plane magnetic field at 4.2~K for \#A, respectively. (c,d) Second
derivative of $\rho_{xx}(B)$ as a function of $1/B$ and $1/B_{\parallel}$ to highlight the
SdH oscillations. (e) Extracted frequencies from the FFT analysis as a function of angle
showing contributions from surface and bulk carriers. The straight (dashed) lines correspond
to the $1/$cos$\Theta$ dependence expected for a purely two-dimensional system. (f,g) FFTs for
a perpendicular and parallel field sweep, respectively, with the primary oscillation frequencies
highlighted.}
\end{figure}

We now turn to focus on the observation of quantum oscillations which is presented in
Fig.~\ref{fig2} for the sample with the highest carrier mobility (\#A).
In Figs.~\ref{fig2}(a,b), we show $\rho_{xx}(B)$ at 4.2~K when subject to a
out-of-plane and in-plane magnetic field. Quantum oscillations are clearly visible
in the second derivative $-d^2\rho_{xx}/dB^2$, respectively, as a function of the inverse field, plotted in
Figs.~\ref{fig2}(c,d) for both orientations. In the parallel field configuration, only one
frequency is evident, whereas several frequencies are found in a perpendicular field.

In order to identify the origin of the quantum oscillations, we have performed Fast Fourier Transforms 
(FFTs) on a series of $\rho_{xx}$ curves measured at different tilt angles $\Theta$. The results
are summarized in Fig.~\ref{fig2}(e). For $\Theta$=0 (perpendicular field configuration),
we observe three frequencies at 21.5, 100 and 162~T (see Fig.~\ref{fig2}(f)). All frequencies 
appear to follow a $1/$cos$\Theta$ up to tilt angles of around 60$^{o}$ characteristic of a 
two-dimensional electronic state. Beyond 60$^{o}$, however, the lower frequency starts to deviate
from this behavior and saturates towards 90$^{o}$. We therefore attribute the observed frequencies 
to two surface states (top and bottom) and one bulk band. Taking the Onsager relation, i.e. the 
extremal cross section of the Fermi surface $A(E_F)\propto f$ and assuming an ellipsoid pocket 
with $V = 4/3\pi a^2b$, we obtain $n_{bulk}$=8.1$\cdot$10$^{17}$~cm$^{-3}$ for
the bulk band corresponding to the pocket with the lowest frequency. For the surface states, we 
obtain the carrier densities 2.4$\cdot$10$^{12}$~cm$^{-2}$ and 3.9$\cdot$10$^{12}$~cm$^{-2}$. 
From the quantum oscillation analysis, we thus obtain a total carrier concentration of 
$n_{tot,SdH}$ = 1.0$\cdot$10$^{18}$~cm$^{-3}$, in excellent agreement with $n_{Hall}$. 
In contrast, assuming that all three pockets were ellipsoidal (i.e. bulk), we would obtain a 
total carrier concentration that is one order of magnitude larger than $n_{Hall}$.

\begin{figure}[ht]
\includegraphics[width=8cm]{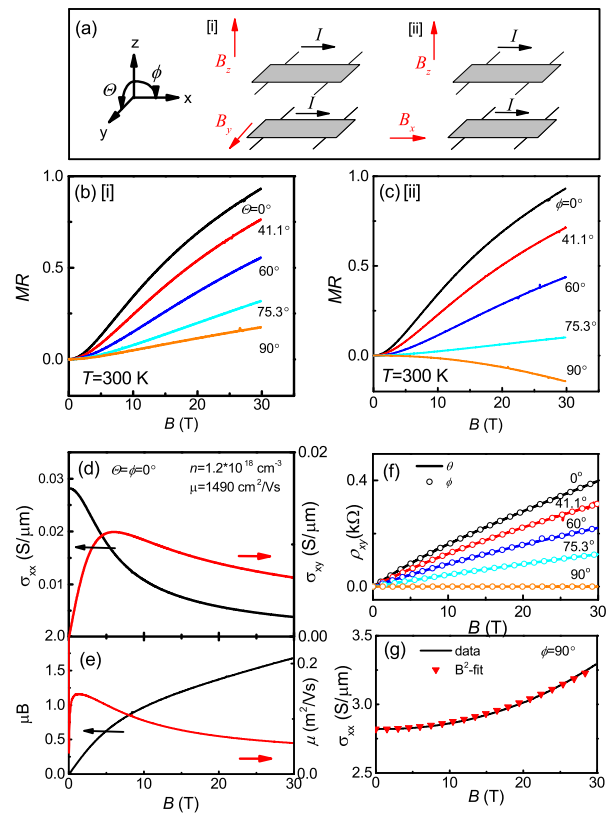}
\caption{\label{fig3} (Color online) Anisotropic magneto-transport in Bi$_2$Se$_3$
at $T$=300~K: (a) Schematic diagram of the electrical transport measurements for
configurations [i] and [ii]. (b,c) Magneto-resistance of sample \#A as a function of $B$ for both
configurations indicating a strong negative MR if $I \parallel B_x$. (d) Longitudinal $\sigma_{xx}$
and Hall conductivities $\sigma_{xy}$ as a function of the magnetic field and (e) extracted
$\mu B$ and $\mu$ using the Drude model. (f) Hall resistivity $\rho_{xy}$ as a function of
$B$ for both configurations (solid line for [i], open symbols for [ii]). (g) The conductivity
$\sigma_{xx}$ as a function of $B$ (solid line) at $\phi = 90^o$ is found to be $\propto B^2$
(triangles).}
\end{figure}

Let us now turn our attention to the peculiar MR we observe in these samples. To avoid quantum oscillatory
and quantum interference phenomena~\cite{29}, we first focus here on the angle-dependent
MR response at room temperature. The longitudinal and Hall resistivities have been measured in 
two different configurations, as shown in Fig.~\ref{fig3}(a). In configuration [i], the applied 
magnetic field is always perpendicular to $I$ (field rotated in the orthogonal plane) whereas 
in configuration [ii], the current and field are parallel if $\phi = 90^o$. The carrier 
concentrations (mobilities) extracted from $\rho_{xy}$ ($\rho_{xx}$) at low fields are 
summarized in Table~1 in the Supplemental material~\cite{29}.

We first present our results and analysis for the high mobility sample (\#A) in Fig.~\ref{fig3}.
In Figs.~\ref{fig3}(b) and (c), we plot the MR=($\rho_{xx}(B)-\rho_0$)$/\rho_0$ at different 
angles $\Theta$ and $\phi$ as indicated in each figure. The overall MR is similar to the one 
observed at low temperature, i.e. it first increases quadratically then tends towards saturation 
at higher field. In both configurations, the MR is strongly anisotropic. Most surprisingly, we 
observe a large NLMR ($\sim$~15~\%) when the magnetic field is applied parallel to the 
current ($I\parallel B_x$). As the second bulk conduction band is far from the Fermi energy $\epsilon_F$ at room 
temperature \cite{30,31,32}, we analyze the MR at $\Theta$=$\phi$=0 using a standard 
one-carrier Drude model (for completeness, a two-carrier analysis is presented in the Supplemental 
Material~\cite{29}).

The corresponding longitudinal and Hall conductivities $\sigma_{xx}$ and $\sigma_{xy}$
in the transverse configuration are illustrated in Fig.~\ref{fig3}(d). From $\rho_{xy}/\rho_{xx}$
we extract $\mu B$ and finally the carrier mobility $\mu$ as a function of the applied field, as shown in
Fig.~\ref{fig3}(e), and find that $\mu$(0~T)/$\mu$(30~T) $\simeq$ 2.7. Based on this simple analysis,
we can infer that the mobility and corresponding scattering time strongly depend on the magnetic field.
We have also measured the dependence of $\rho_{xy}$ for both configurations and found that
the Hall resistivity follows a simple cosine dependence and does not depend on the orientation of $B$ 
with respect to $I$ (see Fig.~\ref{fig3}(f)). Finally, in Fig.~\ref{fig3}(g), we plot $\sigma_{xx}(B)$ 
for the parallel field configuration ($I \parallel B_x$ - solid line) and observe a $B^2$-dependence
up to 30~T (symbols represent a quadratic fit $\sigma_{xx}(B)=\sigma_0+aB^2$ to the data).

\begin{figure}[ht]
\includegraphics[width=8cm]{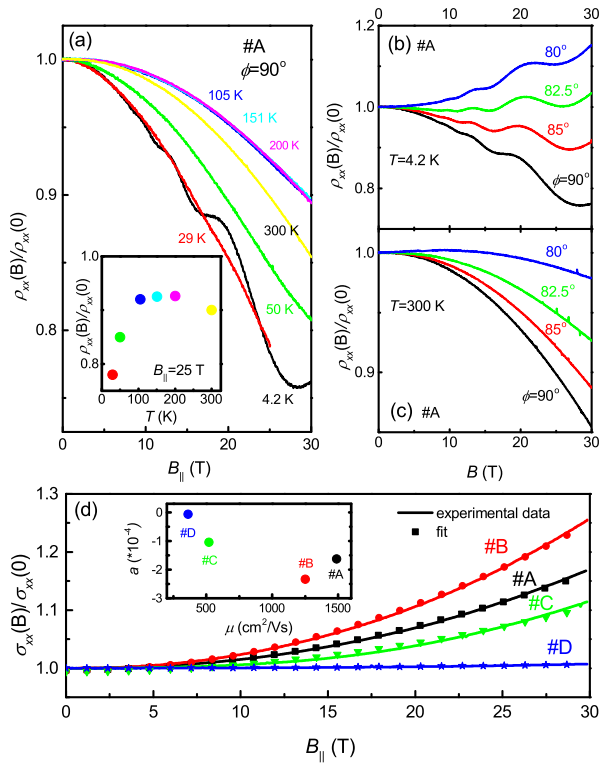} 
\caption{\label{fig4} (Color online) (a) Normalized longitudinal magneto-resistivity $\rho_{xx}(B)/\rho_{xx}(0)$ 
for sample \#A at different temperatures (inset: $\rho_{xx}(B)/\rho_{xx}(0)$ at $B$=25~T). $\rho_{xx}(B)/\rho_{xx}(0)$ 
for different angles $\phi$ at (b) $T$ = 4.2~K (c) and $T$ = 300~K. (d) Normalized magneto-conductivity 
$\sigma_{xx}(B)/\sigma_{xx}(0)$ at $T$ = 300~K for all samples \#A, \#B, \#C, and \#D for $\phi$ = 90$^o$. 
All samples follow a $\sigma_{xx}(B)/\sigma_{xx}(0)\propto B^2$ dependence (see fits). The inset shows the 
fitting parameter $a$ as a function of $\mu$ for all samples.}
\end{figure}

In Fig.~\ref{fig4}(a), we present the temperature dependence of the normalized longitudinal 
magneto-resistivity $\rho_{xx}(B)/\rho_{xx}(0)$ for several chosen temperatures for sample \#A. 
With increasing temperature the NLMR becomes slightly weaker, is constant in the range between 
100 and 200~K and increases again at 300~K as shown in the inset to Fig.~\ref{fig4}(a).
In Figs.~\ref{fig4}(b) and (c), we plot $\rho_{xx}(B)/\rho_{xx}(0)$ for sample \#A at
4.2 and 300~K, respectively, when an additional perpendicular magnetic field is applied. The NLMR
turns into a positive one by adding a small out-of-plane component at $\phi \simeq 83^o$ ($\phi < 80^o$) 
for 4.2~K (300~K). At 4.2~K, the NLMR is superimposed by SdH oscillations which have previously been
attributed to TSS from the sidewalls \cite{33}. We have shown here, however, that they originate from the
lowest bulk conduction band.

The anisotropy in the MR and the large NLMR in a parallel field are not unique to one particular wafer
or sample. Indeed, for all samples, we observe a positive MR in a purely perpendicular magnetic field 
and a NLMR in the longitudinal configuration \cite{29}. Moreover, for the samples (\#C and \#D) with the 
lowest carrier mobility, a negative MR is also observed for $I \perp B_y$. The negative MR for $I \perp B_y$ 
can be explained using the classical Drude model provided the bulk carriers have a low mobility \cite{29}. 
In Fig.~\ref{fig4}(a), we plot $\sigma_{xx}(B)/\sigma_{xx}(0)$ as a function of $B_{\parallel}$ 
and find that the NLMR gets progressively weaker with decreasing $d$ (increasing carrier concentration and 
decreasing carrier mobility) at room temperature. Significantly, the longitudinal conductivity follows the 
$B^2$-behavior for all samples (The fitting parameter $a$ as a function of $\mu$ is shown in the inset of
Fig.~\ref{fig4}(d) for all samples). 

Standard Boltzmann theory does not predict any longitudinal magneto-resistance in the presence 
of a magnetic field that is parallel to the applied electric field. A NLMR has been observed previously 
in both 1D \cite{34} and 2D \cite{35} charge ordered systems for currents applied parallel to the conducting chains 
(planes). In both cases, the NLMR exhibited ($B/T$)$^2$ scaling attributed to a closing of the charge 
gap due to Zeeman splitting. In Bi$_2$Se$_3$, by contrast, there is no strong $T$-dependence 
in the NLMR. A classical origin, found in inhomogeneous conductors and 
attributed to macroscopic inhomogeneities and thus distorted current paths \cite{36} can be excluded 
since the anisotropic MR does not depend on the lateral sample size \cite{37}. The origin of the anisotropy of the 
MR and in particular, the large NLMR  in Bi$_2$Se$_3$ is likely to arise from the underlying scattering 
mechanism, as inferred from our simple Drude analysis. In 1956, Argyres and Adam predicted a NLMR for 
a 3D electron gas in the case of non-degenerate semiconductors where ionized impurity scattering 
is present \cite{38} as observed, for example, in indium antimonide in the extreme quantum 
limit \cite{39}. In contrast, recently triggered by the discovery of new Dirac materials 
\cite{22,23,24,25,26}, it has been proposed that a quantum mechanical phenomenon called the axial 
anomaly can give rise to a NLMR \cite{27}. In a magnetic field, charge carriers are subject to 
Landau quantization with a one-dimensional (1D) dispersion along $B$. If in addition an 
electric field is applied parallel to $B$, a uniform acceleration of the center of mass in this 
field-induced 1D system produces the same axial anomaly effect as charge pumping between Weyl 
points in a Weyl semi-metal and the subsequent charge imbalance leeds to a NLMR \cite{22,27}. This effective
reduction in the dimensionality of the electronic dispersion is also the proposed origin for the recent 
observation of NLMR in the interplanar resistivity of 2D correlated metals \cite{40}. Remarkably,
the appearance of the NLMR is not tied to the band structure of a particular material, but
rather related to the type of scattering mechanism present in the system and as in the classical
model \cite{37}, ionized impurity scattering is proposed to give rise to a positive magneto-conductivity
$\sigma \propto B^2$ \cite{27}. Depending on the dominant contribution of the underlying scattering mechanisms,
the magneto-conductivity may be temperature-dependent as observed in indium antimonide \cite{39}. 
For Bi$_2$Se$_3$, we estimate that the quantum limit is reached at a field strength $B_0\simeq$ 43~T 
for the sample with the lowest carrier concentration of 1.2$\cdot$10$^{18}$cm$^{-3}$ (sample \#A) and 
thus our experiments lie outside the regime where a transition from a negative to a positive MR is 
proposed to occur due to short-range neutral impurity scattering \cite{27}.

In conclusion, we have investigated the MR response of thin Bi$_2$Se$_3$ epilayers. The
low-temperature angle-dependent SdH data suggests a coexistence of bulk and
surface charge carriers. At room temperature, we find a strong positive MR with a
field dependence that can be explained by a field-dependent carrier mobility.
The magnetoresistance itself is strongly anisotropic and depends on the orientation
of the current $I$ with respect to the parallel component of the magnetic field $B$.
We have demonstrated that the observation of a NLMR akin to the axial anomaly is not 
specific to Dirac or Weyl semi-metals, but may in fact occur in generic three-dimensional 
materials.

Part of this work has been supported by EuroMagNET II under the EU
contract number 228043 and by the Stichting Fundamenteel Onderzoek
der Materie (FOM) with financial support from the Nederlandse
Organisatie voor Wetenschappelijk Onderzoek (NWO). S.W. thanks NWO for his
Veni grant (680-47-424) and acknowledges revealing discussions with B. A. Piot and M. Orlita.


\begin{thebibliography}{32}
\bibitem{37}
In addition to the lateral sample sizes (length $L~\times$~width $W$ - (30$\times$10)$~\mu m^2$) 
presented in this paper, we have also measured samples with larger dimensions (600$\times$200)$~\mu m^2$.
\end{thebibliography}
\end{document}